\documentclass[aps,pre,twocolumn,nofootinbib,tightenlines,superscriptaddress,nopacs,amsmath,amssymb,final,letterpaper]{revtex4-1}

\usepackage[utf8]{inputenc}
\usepackage{calc}
\usepackage{graphicx}
\usepackage{amsmath,amssymb,amsthm}
\usepackage{bbold}
\usepackage{bm}
\usepackage{color}
\usepackage{calrsfs}
\usepackage[dvipsnames]{xcolor}
\usepackage{enumitem}
\DeclareMathOperator*{\argmax}{argmax}

\usepackage[colorlinks=true,linkcolor=blue,urlcolor=blue,citecolor=blue,anchorcolor=blue]{hyperref}

\begin{document}
\title{Clustering of heterogeneous populations of networks}

\author{Jean-Gabriel \surname{Young}}
\email{Equal contributors}
\affiliation{Department of Mathematics and Statistics, University of Vermont, Burlington VT, USA}
\affiliation{Vermont Complex Systems Center, University of Vermont, Burlington VT, USA}

\author{Alec \surname{Kirkley}}
\email{Equal contributors}
\affiliation{Department of Physics, University of Michigan, Ann Arbor MI, USA}
\affiliation{School of Data Science, City University of Hong Kong, Hong Kong}

\author{M. E. J. \surname{Newman}}
\affiliation{Department of Physics, University of Michigan, Ann Arbor MI, USA}
\affiliation{Center for the Study of Complex Systems, University of Michigan, Ann Arbor MI, USA}

\date{\today}

\begin{abstract}
Statistical methods for reconstructing networks from repeated measurements typically assume that all measurements are generated from the same underlying network structure.  This need not be the case, however.  People's social networks might be different on weekdays and weekends, for instance.  Brain networks may differ between healthy patients and those with dementia or other conditions.  Here we describe a Bayesian analysis framework for such data that allows for the fact that network measurements may be reflective of multiple possible structures.  We define a finite mixture model of the measurement process and derive a Gibbs sampling procedure that samples exactly from the full posterior distribution of model parameters.  The end result is a clustering of the measured networks into groups with similar structure.  We demonstrate the method on both real and synthetic network populations.
\end{abstract}

\maketitle

\section{Introduction}
Many modern network analyses involve large corpora of networks defined on a constant set of nodes.  Examples include repeated observations of proximity networks~\cite{eagle2009inferring}, longitudinal studies of social relationships among fixed groups of individuals~\cite{butts2003network}, and measurements of neural connectivity across patients~\cite{sporns2010networks}. 

As pointed out in a number of recent studies, repeated measurements offer a unique opportunity to carry out robust network analyses~\cite{fienberg1985statistical,priebe2015statistical,newman2018network,newman2018estimating,peixoto2018reconstructing,le2018estimating,tang2018connectome,wang2019common,young2020bayesian,young2021reconstruction}.  With such large, rich data sets it is possible to extract information not available from a single network observation.  For example, one can use repeated observations to infer a latent network structure despite the presence of measurement error---given multiple noisy realizations of the same network it becomes possible to estimate the network most likely to have generated the measurements.  In the case of repeated observation of social interactions, for instance, one expects friends to interact more often than strangers, and repeated observations, properly analyzed, can thus reveal the network of friendships~\cite{newman2018network}.  In neuroscience applications, one might focus on the identification of a characteristic brain connectivity pattern for a set of patients with a common condition, which is uncovered by combining the measurements made on all patients~\cite{sporns2010networks}.

Implicit in these methods is the assumption that there is a single underlying network that determines all the observations, but this need not be the case.  Patients undergoing brain scans might have different underlying conditions, for instance, leading to different measured connectivity patterns from one patient to another~\cite{bassett2018understanding}.  Likewise, acquaintances might interact differently based on the time of the day or location, leading for example to different social networks during working hours and outside of work.

Treating a population of observed networks as realizations of the same underlying graph might thus adversely affect our analysis of these systems, just as ignoring the multimodal nature of a distribution of numbers may lead to mischaracterization of a study sample.  In some cases we know when the underlying network changes, such as when measurements are made on different days of the week, in which case the data can be divided up and processed separately using standard techniques.  Often, however, the underlying generative processes are unknown, making it hard to split the sample manually.  For example, if the brain activity of a patient is measured as they are put under anesthesia, we know that the activity will undergo a change from one state to another but the transition point itself is unknown.  It could be that the activity changes well before a visible loss of consciousness, or well after, so one cannot rely on direct observations to determine the change point; one needs to infer the transition from the networks.  In this and other similar cases, one must simultaneously determine whether the data is best described by a single underlying network or many, what these networks are, and which measurements should be attributed to which underlying network.

In this paper we describe a framework for modeling heterogeneous populations of networks.  We view each network in a population as a noisy realization of one underlying network out of multiple possibilities, which we call \emph{modes}.  We model the data as being generated from a superposition or finite mixture of graph distributions~\cite{titterington1985statistical,mclachlan2004finite}, which provides sufficient flexibility to accommodate highly heterogeneous populations.  We demonstrate how our methodology allows us to simultaneously infer the underlying networks and cluster the observed networks such that each cluster of observations consists of noisy realizations of a single underlying network mode.  Our framework also provides a natural means for selecting the number of modes that describe a given set of networks.

Previous work on statistical modeling of network measurements has mostly focused on the unimodal case, where it is assumed that a population of networks is best described by a single underlying network, of which the observations are noisy realizations~\cite{fienberg1985statistical,kenny1984social,banks1994metric,butts2003network,newman2018network,peixoto2018reconstructing,lunagomez2020modeling,josephs2021network}.
Some recent approaches have explicitly incorporated multiple modes using latent space representations~\cite{durante2017nonparametric,nielsen2018multiple,wang2019joint,arroyo2019inference}, exponential random graph models~\cite{yin2019finite}, or parametric network models~\cite{signorelli2020model}.  Further afield are studies aiming to cluster the layers of multilayer networks (where layers can be categorical or temporal), for instance when trying to detect change points~\cite{peel2015detecting,peixoto2018change}---abrupt changes in sequences of network snapshots---or as a side-effect of pooling information across network layers when clustering their nodes~\cite{peixoto2015inferring,stanley2016clustering}.  Among these works the approach most closely related to our own is perhaps that of La~Rosa~\textit{et al.}~\cite{la2016gibbs}, who extend unimodal metric models of network populations~\cite{banks1994metric,lunagomez2020modeling} to the multimodal case with finite mixtures.  In the unimodal case all possible networks are assigned a distance to the mode and networks closer to the mode are assumed more likely.  The extension described by La~Rosa~\textit{et al.}\ employs multiple modes and can thus more faithfully model diverse populations of networks.  While mathematically elegant, however, this approach has some shortcomings.  For example, the model is not easily estimated when the distance between networks is difficult to compute~\cite{bento2019family}.
The approach of La~Rosa~\textit{et al.}\ also does not differentiate between false-positive and false-negative rates, which may differ substantially and change the composition of the corresponding clusters of networks (see Appendix~\ref{appendix:metric}).

This paper is organized as follows.  First, we briefly review a previously described unimodal model of homogeneous populations of networks, then introduce our model for multimodal heterogeneous populations by building upon the unimodal case.  We then discuss statistical estimation of the model using a Gibbs sampling procedure and demonstrate using synthetic data that we can recover model parameters when the level of noise in the measurements is sufficiently low.  We further demonstrate our methods with an example application to a real-world network population from a longitudinal social network study.

% ~~~~~~~~~~~~~~~~~~~~~~~~~~~~~~~~
\section{The model}
\label{sec:model}
% ~~~~~~~~~~~~~~~~~~~~~~~~~~~~~~~~

We consider an experiment or observational study in which $N$ networks are measured on the same set of $n$ nodes.  The networks could record, for instance, connectivity patterns in brain scans of a cohort of $N$ patients, in which a node represents a region of the brain and edges indicate when two regions are sufficiently connected in a given patient~\cite{bassett2017network}.  Or the population of networks could encode a set of relationships among a group of people such as students in a school with nodes representing the students and edges indicating when two students are within a certain physical distance of one another during a specified time interval~\cite{stehle2011high}.  We record the networks as a set of $N$ adjacency matrices $\mathcal{D}=\{\bm{D}^{(t)}\}_{t=1}^{N}$ indexed by $t=1,\dots,N$, where $\bm{D}^{(t)}$ is an $n\times n$ matrix with element $D^{(t)}_{ij}=1$ if there is an edge between nodes $i$ and $j$ in network~$t$, and $D^{(t)}_{ij}=0$ otherwise.  For simplicity of presentation we will assume the networks to be undirected so that $D^{(t)}_{ij}=D^{(t)}_{ji}$, but our methods are easily adapted to directed networks.

\subsection{Homogeneous populations of networks }
A variety of approaches have been proposed for analyzing multiple network observations of this kind when each observation is believed to be a noisy realization of a single underlying ``ground truth'' network---see for example \cite{butts2003network,banks1994metric,newman2018network,le2018estimating,peixoto2018reconstructing,lunagomez2020modeling}.  As a starting point for our discussion we adopt the approach of~\cite{butts2003network,newman2018network} in which one defines a model to describe how the observed networks are related to the underlying ground truth.  The particular model in this case has two parameters: $\alpha\in[0,1]$ is the a true-positive rate for the edges and $\beta\in[0,1]$ is the false-positive rate.  In other words, if there is an edge connecting two nodes in the adjacency matrix~$\bm{A}$ of the ground truth network, then that edge will be observed with independent probability~$\alpha$ in a noisy realization~$\bm{D}^{(t)}$, while an absent edge in $\bm{A}$ will be mistakenly observed with probability~$\beta$.  The probability of observing a complete network~$\bm{D}^{(t)}$ under this model is then given by
\begin{multline}
  \label{eq:homogeneous_model}
  P(\bm{D}^{(t)}|\bm{A}\!,\alpha,\beta) = \prod_{i < j} \left[\alpha^{D^{(t)}_{ij}}(1-\alpha)^{1-D^{(t)}_{ij}}\right]^{A_{ij}}\\\times \left[\beta^{D^{(t)}_{ij}}(1-\beta)^{1-D^{(t)}_{ij}}\right]^{1-A_{ij}},
\end{multline}
where the product is over all (unordered) pairs of nodes.
When the individual observed networks in $\mathcal{D}$ are independent of one another the likelihood of the complete population~$\mathcal{D}$ is
\begin{align}
   P&(\mathcal{D}|\bm{A}\!, \alpha,\beta)    \label{eq:homogeneous_population_likelihood}
    = \prod_{t=1}^N P(\bm{D}^{(t)}|\bm{A}\!,\alpha,\beta) \notag\\
   &= \prod_{t=1}^N\prod_{i < j} \left[\alpha^{D_{ij}^{(t)}}(1-\alpha)^{1-D_{ij}^{(t)}}\right]^{A_{ij}} \!\!\left[\beta^{D_{ij}^{(t)}}(1-\beta)^{1-D_{ij}^{(t)}}\right]^{1-A_{ij}} \notag\\
   &= \prod_{i < j}\! \left[\alpha^{X_{ij}}\!(1-\alpha)^{N-X_{ij}}\!\right]^{A_{ij}} \!\left[\beta^{X_{ij}}\!(1-\beta)^{N-X_{ij}}\!\right]^{1-A_{ij}},
\end{align}
where $X_{ij}=\sum_{t}D_{ij}^{(t)}$ is the total number of times an edge is observed between node $i$ and $j$ in the $N$ samples.

With Eq.~\eqref{eq:homogeneous_population_likelihood} in hand, one can simulate observed networks or perform inference about the generative process and, for instance, estimate $\bm{A}$ from the data~$\mathcal{D}$.  Applications of this kind are studied at length in Refs.~\cite{butts2003network,newman2018network,peixoto2018reconstructing,young2020bayesian} among others.  Our goal here is to extend the model to accommodate heterogeneous network populations, making it suitable for inference about a broader range of network data.

\subsection{Heterogeneous populations of networks}
Consider again our sample~$\mathcal{D}$ of $N$ networks sharing the same set of nodes.  We can allow for heterogeneity in these samples by letting the individual networks be noisy realizations of $K\ll N$ different underlying network modes~$\mathcal{A}=(\bm{A}^{(1)},...,\bm{A}^{(K)})$, rather than just a single mode as before.  In the context of network neuroscience, for example, we could repeatedly measure the brain of a single patient $t$ undergoing a transition from the conscious to unconscious state, which would imply that $K=2$.  Alternatively, we might have a population of patients, some of whom have a neurological disorder such as Alzheimer's disease and some of whom do not.

Each sample $t$ will be assigned to a network mode~$u$ so that $\bm{D}^{(t)}$ is a noisy realization of~$\bm{A}^{(u)}$.  We use a variable $z_{tu}$ to encode this information thus:
\begin{equation}
  z_{tu} = \left\{
  \begin{array}{ll}
    1 & \text{if $\bm{D}^{(t)}$ is a noisy realization of network $\bm{A}^{(u)}$,}\\
    0 & \text{otherwise.}
  \end{array}
  \right.
\end{equation}
With this notation we can denote the assignment of samples to modes as a $N\times K$ matrix $\bm{Z}$, whose rows correspond to samples and columns to reference networks.  Each sample corresponds to precisely one reference network, so the rows of $Z$ satisfy $\sum_{u=1}^{K} z_{tu} = 1$.  Further, the column sums $N_u = \sum_{t=1}^{N} z_{tu}$ correspond to the number of samples in $\mathcal{D}$ generated from mode~$u$.

To model the now heterogeneous population of networks we use the same generative process as before, Eq.~\eqref{eq:homogeneous_model}, but with a set~$\mathcal{A}$ of multiple underlying networks instead of just one.  Since different modes may display different rates of measurement error, we allow each mode $u=1,\dots,K$ to have its own associated true- and false-positive rates $\alpha_u$ and~$\beta_u$.  For notational simplicity we henceforth denote the sets of parameters $\{\alpha_u\}_{u=1}^{K}$ and $\{\beta_u\}_{u=1}^{K}$, as well as other model parameters we will introduce shortly, collectively as~$\bm{\theta}$.

The likelihood of this model is analogous to that of Eq.~\eqref{eq:homogeneous_population_likelihood} and is given by
\begin{multline}
  \label{eq:heterogeneous_likelihood}
  P(\mathcal{D}|\bm{Z},\mathcal{A},\bm{\theta})  =
  \\
  \prod_{t=1}^N \prod_{u=1}^K \Biggl[\prod_{i < j} \Bigl[\alpha_u^{D_{ij}^{(t)}}(1-\alpha_u)^{1-D_{ij}^{(t)}}\Bigr]^{A_{ij}^{(u)}} \\ \times \!\!\Bigl[\beta_u^{D_{ij}^{(t)}}(1-\beta_u)^{1-D_{ij}^{(t)}}\Bigr]^{1-A_{ij}^{(u)}} \Biggr]^{z_{tu}}.
\end{multline}
The main difference is the product over modes $u$ and the inclusion of the variable $z_{tu}$ to encode the modes.  Performing the product over network samples $t=1,\dots,N$, we can also write this as
\begin{multline}
  \label{eq:heterogeneous_likelihood_compact}
  P(\mathcal{D}|\bm{Z},\mathcal{A},\bm{\theta})  
  =\\
  \prod_{u=1}^K \prod_{i < j} \left[\alpha_u^{X_{ij}^u}(1-\alpha_u)^{N_u-X_{ij}^u}\right]^{A_{ij}^{(u)}} \\ \times \left[\beta_u^{X_{ij}^u}(1-\beta_u)^{N_u-X_{ij}^u}\right]^{1-A_{ij}^{(u)}},
\end{multline}
where $X_{ij}^u=\sum_{t=1}^N D_{ij}^{(t)} z_{tu}$ is the number of interactions observed between $i$ and $j$ across all noisy realizations of the underlying network~$u$.

In most cases, however, the mode assignments $\bm{Z}$ will be unknown---if not then we could just divide up the the networks into their groups and model them separately as disjoint homogeneous populations.  We can eliminate $\bm{Z}$ by marginalizing over its possible values in Eq.~\eqref{eq:heterogeneous_likelihood} thus:
\begin{align}
  \label{eq:multimodal_marginal}
  P(\mathcal{D}|\mathcal{A},\bm{\theta})  &= \sum_{\bm{Z}} P(\mathcal{D},\bm{Z}|\mathcal{A},\bm{\theta}) \nonumber\\
  &= \sum_{\bm{Z}} P(\mathcal{D}|\bm{Z},\mathcal{A},\bm{\theta}) P(\bm{Z}|\bm{\theta}),
\end{align}
where the sum is over all possible assignment matrices~$\bm{Z}$ and $P(\bm{Z}|\bm{\theta})$ is a prior probability on the assignments.  We choose the convenient categorical prior
\begin{equation}
  \label{eq:multimodal_prior}
 P(\bm{Z}|\bm{\theta}) = \prod_{t=1}^{N} \prod_{u=1}^K \pi_u^{z_{tu}}  \equiv \prod_{u} \pi_u^{N_u},
\end{equation}
with $\pi_u$ the prior probability that a sample $\bm{D}^{(t)}$ is assigned to mode $\bm{A}^{(u)}$ (so that $\sum_{u=1}^{K}\pi_u=1$), and the convention that $\bm{\theta}$ now additionally includes the parameters $\{\pi_u\}_{u=1}^{K}$.  Substituting Eqs.~\eqref{eq:heterogeneous_likelihood} and \eqref{eq:multimodal_prior} into Eq.~\eqref{eq:multimodal_marginal}, we then find that
\begin{equation}
  \label{eq:multimodal_mixture}
  P(\mathcal{D}|\mathcal{A},\bm{\theta})  =
  \prod_{t=1}^N \sum_{u=1}^K \pi_u \alpha_u^{Y^{11}_{tu}}(1-\alpha_u)^{Y^{01}_{tu}}\beta_u^{Y_{tu}^{10}}(1-\beta_u)^{Y_{tu}^{00}}, 
\end{equation}
where
\begin{gather}
Y_{tu}^{00} = \sum_{i<j} (1-D_{ij}^{(t)})(1-A_{ij}^{(u)}), \qquad
Y_{tu}^{11} = \sum_{i<j} D_{ij}^{(t)}A_{ij}^{(u)} \notag\\
Y_{tu}^{10} = \sum_{i<j} D_{ij}^{(t)}(1-A_{ij}^{(u)}), \qquad
Y_{tu}^{01} = \sum_{i<j} (1-D_{ij}^{(t)})A_{ij}^{(u)},
\label{eq:y_matrix}
\end{gather}
which we can think of as the elements of four matrices measuring agreement between samples and modes.  (Refer to Table~\ref{table:summary} in the Appendix for a summary of notations used in this paper.)  For example, $Y_{tu}^{00}$ counts the number of edges that are simultaneously absent in sample~$t$ and mode~$u$ and can be thought of as an entry of a matrix~$\bm{Y}^{00}$ that records such numbers for all modes and samples.

\begin{figure}[!t]
  \centering
  \includegraphics[width=0.93\linewidth]{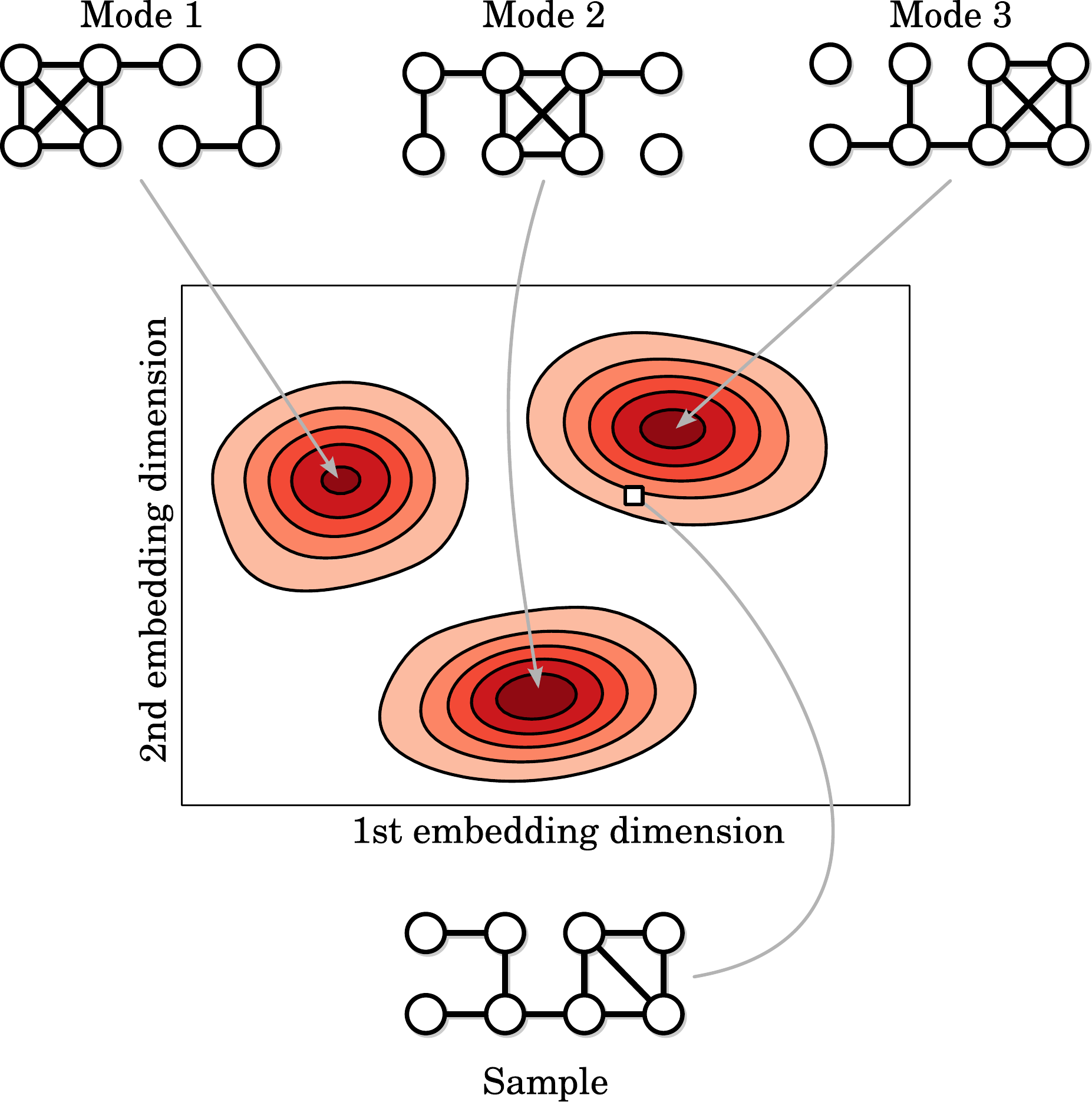}
  \caption{Contour map of a population of networks generated from the heterogeneous model of Eq.~\ref{eq:multimodal_mixture}, as embedded in a two-dimensional space using multidimensional scaling.  Darker colors indicate a higher density of networks and the three peaks correspond to the modes used in the model, which are illustrated at the top.  For this simple example, we draw networks from each of the three modes with identical probability and use the same true- and false- positive rates for all modes, namely $\alpha_u=0.9$ and $\beta_u=0.1$ for $u=1$,~2,~3. As illustrated at the bottom of the figure, a network resembling mode~3 sits close to that mode in the low-dimensional space.}
  \label{fig:MDS}
\end{figure}

The mixture model appearing in Eq.~\eqref{eq:multimodal_mixture} is sufficiently flexible to account for complicated structure in populations of graphs, analogous to the flexibility seen in mixture models over distributions of numbers~\cite{mclachlan2004finite}.  In Fig.~\ref{fig:MDS} we show a population of small networks, generated from Eq.~\eqref{eq:multimodal_mixture} with three modes ($K=3$).  As a visualization of the workings of the model we embed the networks it generates in a two-dimensional space (using multidimensional scaling~\cite{kruskal1978multidimensional} applied to Hamming distance) and then compute the density of network samples in that space, which is shown in the contour plot.  From this plot one can see that the networks generated form clear clusters around the three representative modes, with the model introducing some noise.

% ~~~~~~~~~~~~~~~~~~~~~~~~~~~~~~~~~~~~~~~~~~~~~~~~~~~~~~~~~~~~~~~~~
\section{Estimation of the model using Gibbs Sampling}
\label{sec:algorithm}
% ~~~~~~~~~~~~~~~~~~~~~~~~~~~~~~~~~~~~~~~~~~~~~~~~~~~~~~~~~~~~~~~~~

In a typical application of the model, we observe a population of networks $\mathcal{D}$ and we want to determine $\mathcal{A}$, $\bm{Z}$, and $\bm{\theta}$ assuming that $\mathcal{D}$ was generated from the model as described in Section~\ref{sec:model}.\footnote{We may know some of these parameters in which case the problem can be solved more easily.}
If we can successfully infer the parameters, then we will have not only denoised the network samples~$\mathcal{D}$ by finding each network's mode, but we will have also clustered the data into homogeneous classes of networks---two tasks of considerable scientific interest.  

Here we adopt a Bayesian approach to the estimation problem, which allows us to draw directly from standard model-based clustering techniques \cite{mclachlan2004finite}.  The starting point for this approach is the posterior distribution for the quantities of interest:
\begin{align}
  \label{eq:bayes}
  P(\bm{Z}, \mathcal{A}, \bm{\theta}|\mathcal{D}) =  \frac{P(\mathcal{D}|\bm{Z}, \mathcal{A}, \bm{\theta})P(\bm{Z}, \mathcal{A}, \bm{\theta})}{P(\mathcal{D})},
\end{align}
where $\bm{\theta}$ collectively denotes the rate parameters $\alpha$ and~$\beta$ as well as the mixture weights~$\pi$, and $P(\bm{Z}, \mathcal{A}, \bm{\theta})$ is a prior over all the unknown quantities.  Note that the number~$K$ of reference networks does not appear in the equation; here we use a parametric approach and treat $K$ as a known quantity to be handled separately.  There are a number of more complex Bayesian non-parametric model alternatives one could consider that would permit Gibbs sampling and allow $K$ to vary, including Dirichlet process mixture models \cite{neal2000markov}, but we will not explore these options here.  Instead, the procedure used to infer the posterior distribution in Eq.~\eqref{eq:bayes} can be run for multiple values of~$K$ and the optimal value can be chosen by identifying which value of $K$ produced the samples with the highest probability.  One can add a prior distribution on $K$ to penalize parametrizations with a greater number of clusters, thereby preventing uninformative solutions with $K\simeq N$ from being chosen.  However, the prior probability $P(\bm{Z}|\bm{\theta})$ in Eq.~\eqref{eq:multimodal_prior} already penalizes solutions with $K\gg1$, and this seems sufficient for making reasonable inference in practice---see the results in Sec.~\ref{sec:results}.

We have already discussed the likelihood in Eq.~\eqref{eq:bayes}, which leaves us with the task of specifying the prior distribution $P(\bm{Z},\mathcal{A},\theta)$ to complete the model.  We use a prior that factorizes in the form
\begin{equation}
    P(\bm{Z},\mathcal{A}, \bm{\theta}) = P(\bm{Z}| \bm{\theta})P(\mathcal{A}|\bm{\theta})P(\bm{\theta}).
\end{equation}
The first probability on the right-hand-side, $P(\bm{Z}| \bm{\theta})$, is specified in Eq.~\eqref{eq:multimodal_prior}.  It is the prior over assignments $\bm{Z}$ given the vector of group assignment probabilities~$\pi$, which we have included in~$\bm{\theta}$.  For the other two probabilities we select the simplest priors possible.  For the networks, we set
\begin{align}
\label{eq:priorA}
  P(\mathcal{A}|\bm{\theta}) &= \prod_{u=1}^{K} P(\bm{A}^{(u)}) = \prod_{u=1}^K \prod_{i<j}\rho^{A_{ij}^{(u)}}(1-\rho)^{1-A_{ij}^{(u)}}\notag\\
&= \rho^{M^\ast}(1-\rho)^{K\binom{n}{2}-M^*}
\end{align}
where $M^*=\sum_{i<j}\sum_uA_{ij}^{(u)}$ is the total number of edges in all reference networks and $\rho$, which we also now include in $\bm{\theta}$ for notational simplicity, is the prior probability of an edge being placed between each pair of nodes $(i,j)$.  For the parameters~$\bm{\theta}$---i.e.,~ $\{\alpha_u,\beta_u,\pi_u\}_{u=1}^{K}$ and $\rho$---we use uniform priors so that their overall contribution to the posterior distribution is a multiplicative constant.  (One can instead opt for conjugate prior distributions for these quantities, which would be the beta and Dirichlet distributions in this case, without much mathematical complication, see Appendix~\ref{appendix:priors}.)

With these priors in place, estimation of the heterogeneous network model in Sec.~\ref{sec:model} amounts to either finding point estimates of the parameters in Eq.~\eqref{eq:bayes}, or to computing averages over their distribution.  Here, we employ a sampling algorithm that allows us to accomplish either these tasks. 
The algorithm, described below, returns a series of samples $(\mathcal{A}_s, \bm{Z}_s, \bm{\theta}_s)_{s=1,...,S}$, each giving a possible choice of the network modes, assignment matrices, and parameters from which the networks $\mathcal{D}$ could have been generated.  
These samples can in turn be used to approximate expectations 
\begin{align*}
    E[f(\bm{Z},\mathcal{A},\bm{\theta})] &= \int \sum_{\bm{Z},\mathcal{A}} f(\bm{Z},\mathcal{A},\bm{\theta})  P(\bm{Z}, \mathcal{A}, \bm{\theta}|\mathcal{D}) d\theta \notag \\
    &\simeq \frac{1}{S}\sum_{s=1}^Sf(\bm{Z}_s,\mathcal{A}_s,\bm{\theta}_s),
\end{align*}
of arbitrary functions over the joint posterior distribution of the model, to find estimates of the model parameters, or to assess our uncertainty about the exact values of the parameters.

\subsection{ Gibbs sampling}
We use a Gibbs sampling method to generate our samples~\cite{geman1984stochastic}.  
The Gibbs sampler operates by cycling through the parameters $(\mathcal{A}, \bm{Z}, \bm{\theta})$ of the model and generating values for each set of parameters in turn, while conditioning on the values of the remaining parameters, starting from random initial values.
Specifically, we
\begin{enumerate}
    \item Initialize an iteration counter $\tau\leftarrow 0$ then draw random initial values $\mathcal{A}_0$, $\bm{Z}_0$, $\bm{\theta}_0$ of the parameters.
    \item Draw new modes $\mathcal{A}_{\tau+1} \sim P(\mathcal{A} | \mathcal{D}, \bm{Z}_{\tau}, \bm{\theta}_{\tau}).$
    \item Draw new assignments $\bm{Z}_{\tau+1} \sim P(\bm{Z} | \mathcal{D}, \mathcal{A}_{\tau+1}, \bm{\theta}_{\tau}).$
    \item Draw new parameters $\bm{\theta}_{\tau+1} \sim P(\bm{\theta} | \mathcal{D}, \mathcal{A}_{\tau+1}, \bm{Z}_{\tau+1}).$
    \item Increment $\tau$ and repeat from step~2.
\end{enumerate}%
\noindent Samples drawn from this random process at sufficiently large intervals will be distributed according to the joint posterior distribution of Eq.~\eqref{eq:bayes}~\cite{geman1984stochastic}.

The conditional probability distributions needed for Gibbs sampling can be calculated from the model's likelihood given in Eqs.~\eqref{eq:heterogeneous_likelihood} and~\eqref{eq:heterogeneous_likelihood_compact} and from the priors of Eqs.~\eqref{eq:multimodal_prior} and~\eqref{eq:priorA}.  First, we have the probability of a given set of reference networks $\mathcal{A}$, conditioned on the data $\mathcal{D}$ and the values of $\bm{Z}$ and~$\bm{\theta}$:
\begin{equation}
  \label{eq:reference_update_formal}
  P(\mathcal{A}|\mathcal{D},\bm{Z},\bm{\theta}) = \frac{P(\mathcal{D}|\bm{Z},\mathcal{A},\bm{\theta})P(\mathcal{A}|\bm{\theta})}{\sum_{\mathcal{A}} P(\mathcal{D}|\bm{Z},\mathcal{A},\bm{\theta})P(\mathcal{A}|\bm{\theta})},
\end{equation}
which we can transform into an explicit expression by substituting in the compact likelihood of Eq.~\eqref{eq:heterogeneous_likelihood_compact} and our prior over $\mathcal{A}$.  We find
\begin{equation}
  \label{eq:reference_update_explicit}
  P(\mathcal{A}|\mathcal{D},\bm{Z},\bm{\theta}) = \prod_{u=1}^K \prod_{i < j}  (Q_{ij}^u)^{A_{ij}^{(u)}}(1-Q_{ij}^u)^{1-A_{ij}^{(u)}},
\end{equation}
where 
\begin{equation}
  \label{eq:reference_update_prob}
  Q_{ij}^u = \left[1 + \frac{(1-\rho)}{\rho} \left(\frac{\beta_u}{\alpha_u}\right)^{X_{ij}^u} \left(\frac{1-\beta_u}{1-\alpha_u}\right)^{N_u-X_{ij}^u}  \right]^{-1}
\end{equation}
is the probability that nodes $i$ and $j$ are connected in the $u$th reference network, when we know the cluster assignments and parameter values.

Likewise, the conditional probability of the cluster assignments $\bm{Z}$ can be calculated as
\begin{equation}
  \label{eq:assignement_update_formal}
  P(\bm{Z}|\mathcal{D},\mathcal{A},\bm{\theta}) = \frac{P(\mathcal{D}|\bm{Z},\mathcal{A},\bm{\theta})P(\bm{Z}|\bm{\theta})}{\sum_{\bm{Z}} P(\mathcal{D}|\bm{Z},\mathcal{A},\bm{\theta})P(\bm{Z}|\bm{\theta})}.
\end{equation}
We already have an expression for the denominator---it is the mixture likelihood of
Eq.~\eqref{eq:multimodal_mixture}---and the numerator can be calculated by combining the prior in Eq.~\eqref{eq:multimodal_prior} with another form of Eq.~\eqref{eq:heterogeneous_likelihood},
\begin{align}
 P&(\mathcal{D}|\bm{Z},\mathcal{A},\bm{\theta}) =\nonumber\\
 &\prod_{t=1}^N \prod_{u=1}^K \Bigl[ (\alpha_u)^{Y_{tu}^{11}}(1-\alpha_u)^{Y_{tu}^{01}}  (\beta_u)^{Y_{tu}^{10}}(1-\beta_u)^{Y_{tu}^{00}} \Bigr]^{z_{tu}},
\end{align}
where the four $\bm{Y}$ matrices are the ones defined in Eq.~\eqref{eq:y_matrix}.  Putting everything together we find that
\begin{equation}
  \label{eq:assignement_update_explicit}
  P(\bm{Z}|\mathcal{D},\mathcal{A},\bm{\theta}) = \prod_{t=1}^{N} \prod_{u=1}^{K} R_{tu}^{z_{tu}}
\end{equation}
where
\begin{align}
  \label{eq:assignement_update_prob}
  &R_{tu} = \frac{\pi_u (\alpha_u)^{Y_{tu}^{11}}(1-\alpha_u)^{Y_{tu}^{01}}  (\beta_u)^{Y_{tu}^{10}}(1-\beta_u)^{Y_{tu}^{00}}}{\sum\limits_{v} \pi_v (\alpha_v)^{Y_{tv}^{11}}(1-\alpha_v)^{Y_{tv}^{01}}  (\beta_v)^{Y_{tv}^{10}}(1-\beta_v)^{Y_{tv}^{00}}}\notag
\end{align}
can be interpreted as the probability that sample $t$ is a noisy version of mode $u$, conditioned on known values for the modes and parameters.

The update equation for the remaining parameters is given by
\begin{equation}
  \label{eq:parameter_update_formal}
  P(\bm{\theta}|\mathcal{D},\bm{Z},\mathcal{A}) = \frac{P(\mathcal{D}|\bm{Z},\mathcal{A},\bm{\theta})P(\bm{Z},\mathcal{A}, \bm{\theta})}{\int P(\mathcal{D}|\bm{Z},\mathcal{A},\bm{\theta})P(\bm{Z},\mathcal{A}, \bm{\theta}) d\bm{\theta}}.
\end{equation}
The integral appearing in the denominator has a closed-form solution, but we don't actually need to calculate it since our goal is to sample from the distribution, so we only need to know how Eq.~\eqref{eq:parameter_update_formal} depends on $\bm{\theta}$, and the denominator is, by definition, not a function of~$\bm{\theta}$.  Upon substituting the likelihood and priors into Eq.~\eqref{eq:parameter_update_formal} we find that the conditional distribution for $\bm{\theta}$ factorizes as
\begin{align}
  \label{eq:parameter_update_explicit}
   P(\bm{\theta}|\mathcal{D},&\bm{Z},\mathcal{A}) 
   \propto \rho^{M^*} (1-\rho)^{K\binom{N}{2} -M^*}\prod_{u=1}^K \pi_u^{N_u} 
   \nonumber\\ 
   &\times \prod_{u=1}^K (\alpha_u)^{W_u^{11}}(1-\alpha_u)^{W_u^{01}} 
   (\beta_u)^{W_u^{10}}(1-\beta_u)^{W_u^{00}},
\end{align}
where the matrices
\begin{align}
W_{u}^{pq}=\sum_{t=1}^{N}Y_{tu}^{pq}z_{tu}     
\end{align}
quantify the edge agreement between the mode $u$ and all the network samples in its corresponding cluster (see Table~\ref{table:summary} for a summary).

Equations \eqref{eq:reference_update_explicit}, \eqref{eq:assignement_update_explicit}, and \eqref{eq:parameter_update_explicit} provide us with all the conditional distributions we need to implement the Gibbs sampler.  As mentioned previously, we cycle through the variables $\mathcal{A}$, $\bm{Z}$ and $\bm{\theta}$, generating new samples of each from Eqs.~\eqref{eq:reference_update_explicit}, \eqref{eq:assignement_update_explicit}, and~\eqref{eq:parameter_update_explicit} respectively while using the most recent samples of the other parameters as input.

The conditional distributions all allow for straightforward sampling.  As Eq.~\eqref{eq:reference_update_explicit} and~\eqref{eq:assignement_update_explicit} show, the edges of the modes and the cluster assignments are determined by independent categorical random variables.  Equation~\eqref{eq:parameter_update_explicit} further shows that the parameters $\bm{\theta}$ are independent from one another (when we condition on $\bm{Z}$ and~$\mathcal{A}$), and that they are governed by either a beta distribution ($\alpha$, $\beta$ and $\rho$) or a Dirichlet distribution~($\pi$).  Hence, generating updated parameters for the Gibbs sampler is straightforward since all variables can be simulated with standard univariate sampling methods available in most statistical software packages.

\subsection{Implementation}
\label{subsec:implementation}
The Gibbs sampling approach allows us to sample from the full posterior distribution of Eq.~\eqref{eq:bayes}, which is sufficient for estimating any expectation value of interest.  In its naive form, however, the Gibbs sampler is quite slow.  For example generating new modes takes $O\bigl(K\binom{n}{2}\bigr)$ steps which can become an issue for larger networks.  Fortunately there are some computational tricks we can use to speed up the calculation substantially.

The first observation we make is that the variables~$\bm{Y}$ are related, such that they need not be all computed every time the modes are updated.  Using Eqs.~\eqref{eq:y_matrix} one can show that
\begin{align}
   Y_{tu}^{10} & = M_t - Y^{11}_{tu}, \\
   Y_{tu}^{01} & = M^\ast_u - Y^{11}_{tu}, \\
   Y_{tu}^{00} &= \binom{n}{2} - Y_{tu}^{10} - Y_{tu}^{01} - Y_{tu}^{11},
\end{align}
where $M_u^\ast$ is the number of edges in network mode~$u$, and $M_t$ is the number of edges in the sample~$t$.  By pre-computing $M_t$ from the data $\mathcal{D}$ once and using the above relations we can recover all the matrices~$\bm{Y}$ in terms of the edge agreements~$\bm{Y}^{11}$ and the counts~$M^\ast_u$.  Furthermore, these latter quantities can be computed rapidly by traversing edge lists, which can be done in order $O(n)$ time for sparse networks with $O(n)$ edges, using for example hash tables to store the lists.  Thus, updates of $\bm{Y}$ only take linear time.

A second set of observations allows us to accelerate the calculation by removing redundancies from the processing of network modes.  First, we notice that, conditioned on the modes, the edges of~$\mathcal{A}$ are independent identically distributed Bernoulli variables as shown in Eq.~\eqref{eq:reference_update_explicit}.  Second, we observe that the probabilities $Q^u_{ij}$ needed to generate these edges are identical for all node pairs~$(i,j)$ that occur a given number of times~$X^{u}_{ij}$ in networks belonging to cluster~$u$, as shown by Eq.~\eqref{eq:reference_update_prob}.  Thus, the fundamental quantity needed to track and update the modes is, in effect, $X^u_{ij}$~rather than the modes themselves.

To make use of these observations, we denote the set of edges that occur exactly $l$ times in cluster $u$ as
\begin{align}
T^u_{l}\equiv \{(i,j):X^u_{ij}=l\},    
\end{align}
and the number of unique values of $l$ observed in cluster~$u$ for the current Gibbs sample as~$L_u$.  For each cluster~$u$, we can then replace the $\binom{n}{2}$ Bernoulli trials needed to generate a sample by a fast two-step process.  First, for each of the unique $L_u$ values of $X_{ij}^u$, we generate the number $E^u_l$ of edges that will appear in the mode out of $\vert T_l^u\vert$ independent trials with a probability of success of $Q^u_{l}$ corresponding to the value of $Q^u_{ij}$ for the edge $(i,j)\in T^u_l$.  Then we choose $E^u_l$ edges uniformly at random from the set $T^u_l$ and add these edges to the mode adjacency matrix~$\bm{A}^{(u)}$, which we store using an edge list implemented with a hash table.  We repeat this sampling procedure for all values $l\in L_u$ in the current cluster sample and repeat for all clusters~$u$.

For a given cluster~$u$, this two-step process can be carried out in $\sum_{l\in L_u} (1+\vert T_l^u\vert Q^u_{l})$ operations on average, which equals $K{n\choose 2}$ only in the worst case when every edge occurs a different number of times~$l$ in every cluster.  For a more typical population of $N\gg 1$ sparse networks with $O(n)$ edges, most pairs of nodes~$i,j$ will never be connected in any of the samples, and thus the sets $T_0^u$ are typically much larger than the sets~$T_l^u$ for $l\geq 1$, which will have about $n$ edges.  As a result our alternative sampling scheme offers a drastic improvement in computational complexity, since we do not need to sample all $O(n^2)$ potential edge pairs in~$T_0^u$ and the low value of~$Q^u_{0}$ ensures that we will rarely sample many edges from this set.

With these implementation tricks, we are able to generate complete samples relatively quickly.
For example, for the collection of $418$ networks on $96$ nodes analyzed in Sec.~\ref{subsec:soc_net} below, our implementation (which is written in Python) takes 83 milliseconds on average to generate a sample on a single 2.0 GHz processor, and a thousand samples usually suffice for computing accurate estimates.
 
\section{Results}
\label{sec:results}
In this section we demonstrate the performance of our method using both synthetic (computer-generated) and real-world data sets.

\subsection{Synthetic data}
As a first demonstration of our approach and the proposed model, we test its ability to infer the parameters of a population of networks generated using the model itself.  We use a two-mode configuration with the modes 1 and 3 shown in Fig.~\ref{fig:MDS} as the ground-truth, and a symmetric population of networks with $\pi_1=\pi_2=\frac12$, such that about half of the networks in the sample are noisy samples of mode~1 and the other half are samples of mode~3.  We control the difficulty of the task by varying the true-positive rate $\alpha_u$ and the false-positive rate $\beta_u$, with
\begin{align}
\label{eq:p}
\beta_u=1-\alpha_u=p\leq 0.5    
\end{align}
identical for all clusters $u$. The parameter $p$ is the probability that $D^{(t)}_{ij}$ is a false positive or a false negative---in other words, the probability of flipping an edge/non-edge in~$\bm{A}^{(u)}$.  For each experiment we generate a population of $N=20$, 50, 100, or~200 networks, which we feed into our algorithm as input~$\mathcal{D}$.

\begin{figure}
  \centering
  \includegraphics[width=0.75\linewidth]{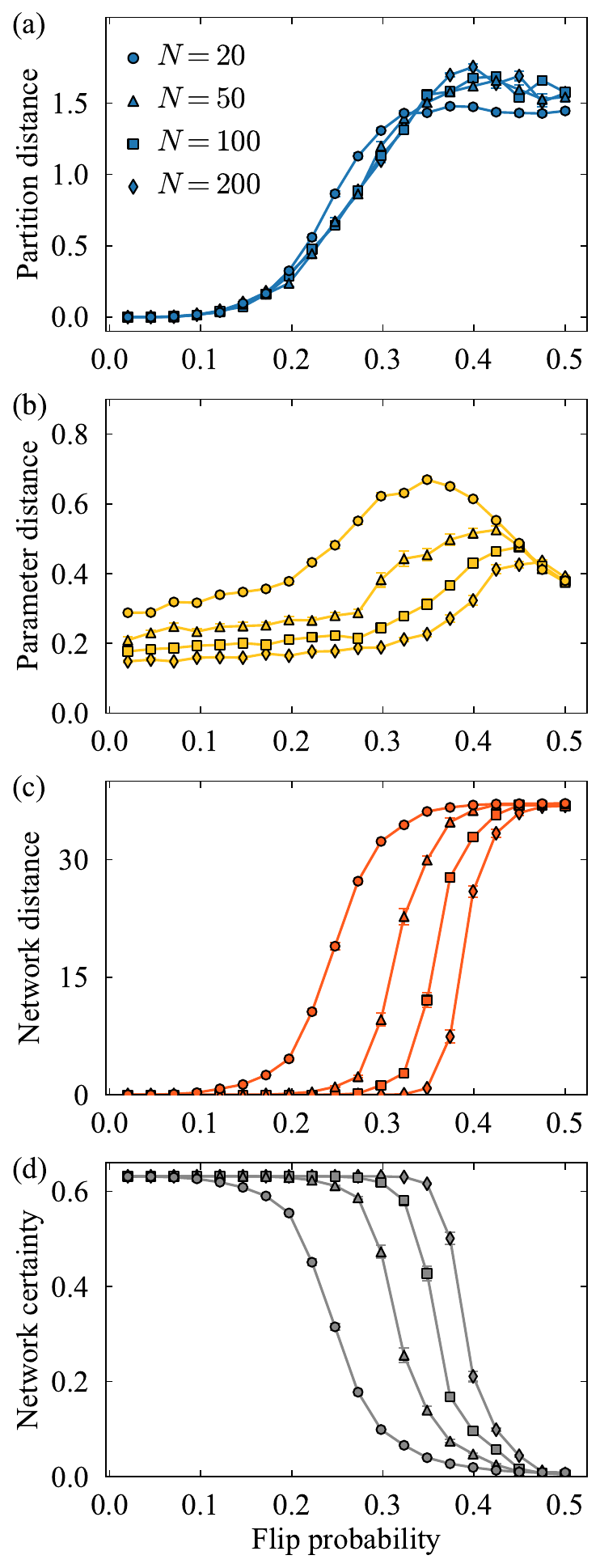} 
  \caption{Recovery performance for (a)~the partitions~$\bm{Z}$, (b)~the model parameters~$\bm{\theta}$, and (c)~the modes~$\mathcal{A}$, for a bimodal population of networks drawn from the modes~1 and 3 shown in Fig.~\ref{fig:MDS} with $\pi_1=\pi_2=\frac12$, across a range of flip probabilities (Eq.~\ref{eq:p}) and population sizes $N$. Panel (d) shows how much the model updates its certainty about the modes $\mathcal{A}$ using the data.}
  \label{fig:synthetic}
\end{figure}

For every synthetic data set~$\mathcal{D}$ we generate $S=5\,000$ samples using the Gibbs sampling algorithm and from these samples compute point-estimates $\{\bm{\hat Z},\mathcal{\hat A},\bm{\hat\theta}\}$. 
For the modes $\hat{A}=(\hat{\bm{A}}^{(1)}, ...\hat{\bm{A}}^{(K)})$ we compute the posterior probability that they exist
\begin{align}
\hat{\bm{A}}_{ij}^{(u)} &= \frac{1}{S}\sum_{s=1}^{S}[A_{ij}^{(u)}]_s
\end{align}
where $[A_{ij}^{(u)}]_s$ is one sample of edge $(i,j)$ of mode $u$,
while for the parameters we use the posterior means directly,
\begin{align}
\bm{\hat\theta} &= \frac{1}{S}\sum_{s=1}^{S} \bm{\theta}^{(s)}.
\end{align}
 We summarize the cluster labels $\bm{\hat Z}$ as a cluster assignment $\bm{\hat g} = \{g_t\}$ such that
\begin{align}
\hat g_t = \argmax_{u}\Big\{\sum_{s=1}^{S}\bm{Z}_{tu}^{(s)}:u\in [K]\Big\}, \end{align}
where $\hat g_t$ is the cluster label of network~$t$ that appears most frequently in the posterior marginal distribution.

Once we have these point estimates we quantify the quality of the reconstruction by comparing them against $\{\bm{Z}_{\text{true}},\mathcal{A}_{\text{true}},\bm{\theta}_{\text{true}}\}$, the parameters used to generate the synthetic networks in the first place.  For the cluster labels~$\bm{g}$ we compute the variation of information~\cite{meilua2007comparing} between the estimates and the true values. For the parameters we compute the $\ell_1$ distance between the parameter vectors $\bm{\theta}$ and~$\bm{\theta}_{\text{true}}$.  And for 
 the modes~$\mathcal{A}$ we compute the total number of missing and spurious edges across all modes (in other words the sum of the Hamming or $\ell_1$ distances between the graphs in $\hat{\mathcal{A}}$ and~$\mathcal{A}_{\text{true}}$).  

We vary the sizes of our synthetic populations over $N=20, 50, 100, 200$ to test how population size affects our reconstruction performance.  For each size we generate 100 realizations of the $N$ networks and take the average reconstruction quality over these synthetic populations (except for $N=20$ where we use 500 realizations to smooth out ineherent variations).  The standard errors associated with the average reconstruction performances are shown as error bars in the plots.  We also use a beta prior for the density $\rho$ with $a^\ast=1$, $b^\ast=20$ (see Appendix \ref{appendix:priors} for details), which allows for more consistent estimates of the mode edge densities in the high noise region where $p\simeq0.5$.

The results of these experiments are shown in Fig.~\ref{fig:synthetic}.  In panel~(a) we show the reconstruction performance for the cluster assignments $\bm{Z}$, which degrades gradually as we increase the flip probability $p$ and does not depend strongly on the number of networks $N$ in the population.  In panel~(b) we show the analogous curves for the parameters $\bm{\theta}$, which also become gradually more difficult to recover as we increase the noise level, up to a certain point, but then become \emph{easier} to recover.  The reason for this is that in the completely noisy limit ($p=0.5$), all the information about the mode networks and clusters is destroyed, and so we end up with completely randomized clusters and modes.  But this means we are likely to end up with values of $\alpha_u$ and $\beta_u$ near the correct value of~0.5, since this is the value for clusters with no structure.  We are also likely to infer values of the~$\pi_u$ close to the correct value of~0.5, as this is the most likely size distribution of the clusters if they are chosen completely at random.  As we approach this regime, therefore, we begin to see improvement in the performance for~$\bm{\theta}$ due to these effects.

Of all the model variables, the modes~$\mathcal{A}$ are the easiest to recover for low levels of noise, as demonstrated by the relatively long flat portions of the curves in panel~(c).  In this case, the modes are recovered near-perfectly for flip probabilities less than some transitional value which depends on~$N$.  Beyond this point the noise introduced into the model through the flip probability~$p$ begins to blur the clusters of networks and introduces errors in their recovery.  As demonstrated by the ordering of the curves in all panels, the reconstruction generally becomes easier as we increase~$N$, which is expected since each new network sample gives us more data about the latent structure.  Reconstructing the partition becomes slightly more difficult for larger $N$ since this gives us more opportunities for making errors.  This effect is reflected in the small increase in partition distance in the regime near $p=0.5$.

When there is little evidence for any specific modes in the data the posterior distribution over modes will be similar to the prior distribution.  Hence, we can quantify the algorithm's certainty about the results using the Kullback-Leibler (KL) divergence between the prior and posterior distributions
\begin{equation}
  D = \sum_{u=1}^K \sum_{i<j} \left[Q_{ij}^u\log \frac{Q_{ij}^u}{\rho} + (1 - Q_{ij}^u)\log \frac{(1 - Q_{ij}^u)}{(1-\rho)}\right],
\end{equation}
where $Q_{ij}^u$ and $\rho$ can be evaluated by averaging the posterior samples.  This divergence is smallest when the posterior and prior distributions are most similar.  As can be seen in Fig.~\ref{fig:synthetic}d, the KL divergence diminishes rapidly as we increase the flip probability and reaches zero at values of $p$ that closely align with the loss of signal seen in panel~(c).  Thus, the method is able not only to recover modes in the low-flip-probability regime, but is also able to tell us that there is no signal when $p$ is large.  Note that the true values of the parameters are not needed for computing the KL divergence, so this approach is applicable in experimental settings where the modes are unknown.

Taken together, the results of Fig.~\ref{fig:synthetic} tell us that our Gibbs sampler is capable of correctly inferring all of the model parameters so long as there is well-defined cluster structure in the data (something we can determine empirically with the KL divergence), and that inference becomes more reliable as the size of the network population grows.  In the following section we demonstrate that our estimation procedure can also successfully cluster populations of real networks and identify multiple distinct modes under real-world conditions.

\begin{figure*}
  \centering
  \includegraphics[width=1\textwidth]{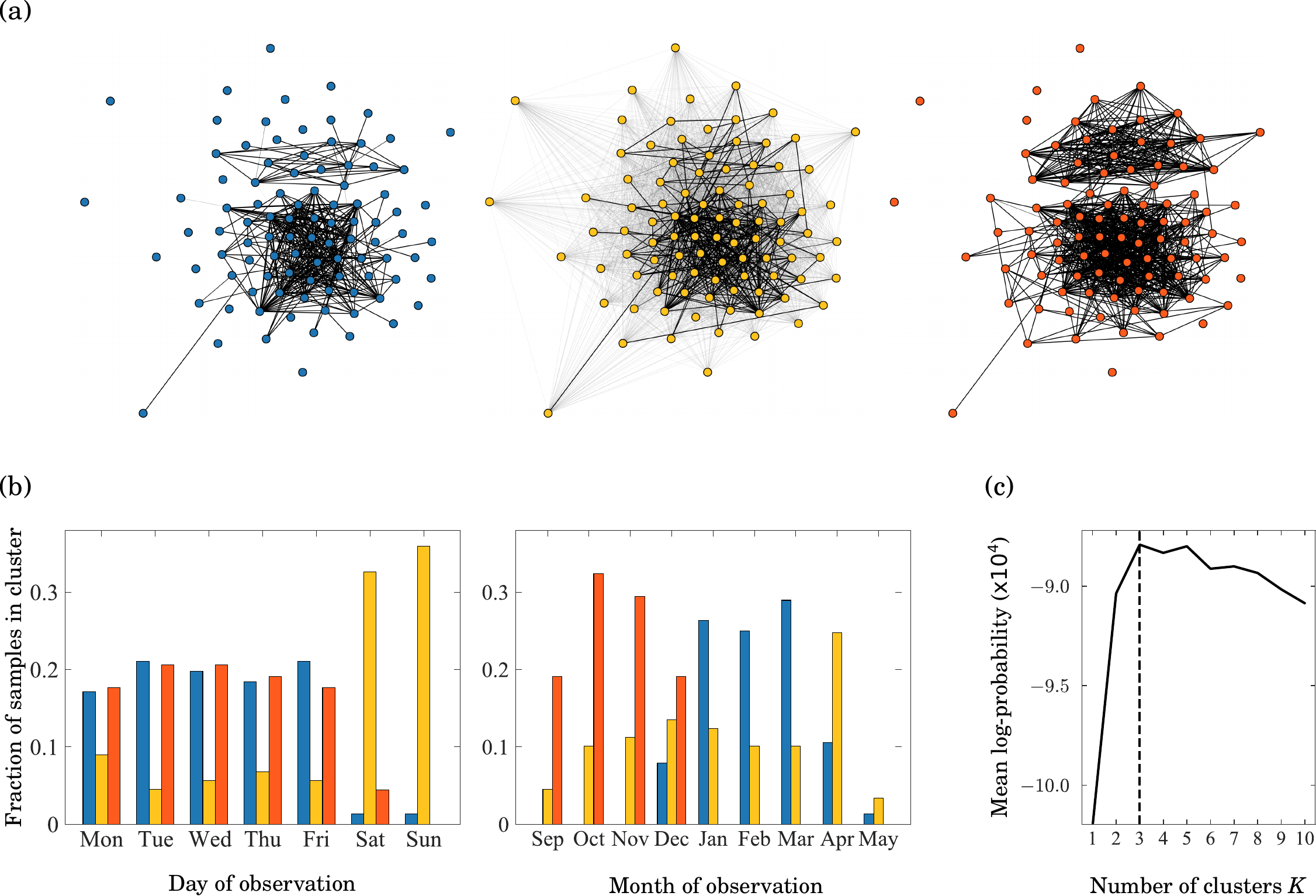}
  \caption{Results for the reality mining proximity networks described in the text, in which proximity measurements were binned in day-long intervals.  (a)~Modal networks~$\mathcal{\hat A}$ with edge widths proportional to the estimates~$\hat A^{(u)}_{ij}$.  (b)~The fraction of networks in each cluster that were sampled on a particular day of the week (left) and during a particular month (right).  (c)~Average posterior log-probability (Eq.~\ref{eq:bayes}) across Gibbs samples as a function of the number of clusters~$K$.}
  \label{fig:real}
\end{figure*}

\subsection{Social network}
\label{subsec:soc_net}
As our second example we study a network of physical proximity measurements among a group of college students and faculty in the ``reality mining'' study of Eagle and Pentland~\cite{eagle2006reality}.  Over a nine-month period, study participants carried mobile phones equipped with special software that recorded when they were in close proximity with one another using Bluetooth radio technology.  From these data we construct networks in which an edge between two participants indicates that they were observed in proximity at least once on a given day.  The result is one network for each of the 232 days of the study where at least one interaction happened, all with n = 96 nodes, and between 1 and 418 edges.

The results of applying our Gibbs sampling algorithm with $S=1000$ samples to this population of networks are shown in Fig.~\ref{fig:real}.  To select the number of clusters~$K$, we run the algorithm at multiple values of~$K$, identifying the value for which the average posterior probability is maximized as the optimal~$K$. In panel~(a) of the figure we display the three very distinct modes~$\mathcal{\hat A}$ found for this population, with the edge widths proportional to the inferred values of~$\hat A^{(u)}_{ij}$.  In panel~(b) we show histograms of the fractions of networks in each cluster that were sampled on a given day of the week (left) and during a given month (right).  We can see clear separation in these histograms, indicating that the modes correspond roughly to weekends (yellow), weekdays during the fall semester (orange), and weekdays during the spring semester (blue).  We see a clearly distinct structure in the networks during semester weekdays versus weekends, with a denser network observed in the first semester than the second.  On the other hand, the weekend network mode contains a wider variety of edges with non-negligible weight.  These observations are consistent with a typical university course schedule in which students associate primarily with classmates during classes but with a wider selection of acquaintances outside of work.  In panel~(c), we show the mean log-probability of the Gibbs samples as a function of the number of modes~$K$, which can we see has a peak at $K=3$ (the value chosen for our posterior estimates).

We note that the modes have significantly higher density than the individual observed networks with their corresponding clusters, as demonstrated by the high false~negative rates $1-\alpha_u$ and low false-positive rates $\beta_u$.  This indicates that the set of likely edges in each mode is much larger than the number of interactions observed in any single time window.  This is not surprising---most people don't see all of their acquaintances every day.  Nonetheless, it emphasizes the ability of our method to infer a plausible set of ``true'' connections from very noisy individual network measurements, a task that would not be easy to do by clustering networks in a naive way.

\section{Conclusions}
In this paper we have described a method for statistical analysis of heterogeneous network populations, as seen in applications involving repeated or longitudinal measurements of a single network, or observation of a fixed set of nodes across multiple study subjects.  We have proposed a generative model for heterogeneous networks in which networks divide into a number of clusters or modes and derived an exact Gibbs sampling procedure for sampling the resulting posterior distributions.  We have demonstrated our model and estimation methods first on synthetic network data, finding that parameter recovery with the Gibbs sampler is possible as long as variability within the clusters of networks is not too high.  We then analyze a real-world network population from a longitudinal proximity network study and find that this population is best described by three underlying mode networks rather than a single ground-truth network as is assumed in most network reconstruction approaches.  Our model is capable of encompassing a greater variety of network data than such unimodal approaches and provides a natural framework for simultaneously clustering and denoising a set of networks.

The framework we describe could be extended in a number of ways.  For the sake of simplicity we have here assumed that our networks are undirected, unweighted, and that that all edges in a given mode have identical error rates.  There are other possibilities, however, including adaptations for directed or weighted edges, more complex noise models such as ones with individual error rates for each node or edge, or the use of a parametric model for the mode networks to allow for simultaneous inference of large-scale structure such as communities.  Additionally, one could employ fast algorithms to fit the model without any sampling whatsoever, by inferring point estimates using methods such as $k$-means clustering after removing the dependence on $\bm{\theta}$ through profiling or marginalization. We leave exploration of these avenues for future work.

\section*{Acknowledgments}
\vspace{-\baselineskip}
We thank Guillaume St-Onge for helpful comments.  This work was funded in part by the James S. McDonnell Foundation (JGY), the National Institutes of Health Centers of Biomedical Research Excellence Award 1P20 GM125498--01 (JGY), the US Department of Defense NDSEG fellowship program (AK), and the US National Science Foundation grant DMS--2005899 (MEJN).

After this work was completed we learned of recent related work by Mantziou~\textit{et al.}  A preprint can be found at~\cite{mantziou2021}.

\appendix

\section{Metric inference}
\label{appendix:metric}
One approach to the problem of network clustering and denoising is to employ methods of \emph{metric inference}~\cite{banks1994metric}, in which network samples live in a metric space and are clustered based on their distance in that space.  It turns out that the simplest metric model of graphs, as used in Refs.~\cite{banks1994metric,lunagomez2020modeling,la2016gibbs}, corresponds to a special case of the model analyzed in the present work.  Here we demonstrate this correspondence.

Starting with the unimodal case, the likelihood of a single network sample under a generic metric model is given by~\cite{banks1994metric}
\begin{equation}
  P(\bm{D}^{(t)}|\bm{A},\sigma) \propto f_\sigma\bigl(d(\bm{D}^{(t)}, \bm{A})\bigr),
\end{equation}
where $\sigma$ is a parameter that controls an analog of the variance, $f_\sigma(\cdot)$ is a non-increasing function, and $d(\cdot,\cdot)$ is a distance metric on the space of all labeled graphs on $n$ nodes.  This model is very flexible in principle but less so in practice, because computational issues place constraints on the possible choices of metric and the function~$f_\sigma$.  Most authors use an exponential function $f_\sigma(x)= e^{-x/\sigma}$ and the Hamming distance~\cite{banks1994metric,lunagomez2020modeling}
\begin{equation*}
  d_H(\bm{D}^{(t)}, \bm{A}) = \sum_{i < j}\left[(1-D_{ij}^{(t)})A_{ij} +D_{ij}^{(t)} (1-A_{ij})\right].
\end{equation*}
One can then use the properties of the exponential to derive the simple data likelihood
\begin{align}
  P(\mathcal{D}|\bm{A}, &\sigma) \notag\\
  &= \prod_{t=1}^N \prod_{i < j}   \left[\frac{e^{-\frac{1}{\sigma} D_{ij}^{t}}}{1+e^{-\frac{1}{\sigma}}}\right]^{(1 - A_{ij})} \left[\frac{e^{-\frac{1}{\sigma} (1-D_{ij}^{t})}}{1+e^{-\frac{1}{\sigma}}}\right]^{ A_{ij}} \notag\\
  &= \prod_{i < j}   \left[\frac{e^{-\frac{1}{\sigma} X_{ij}}}{(1+e^{-\frac{1}{\sigma}})^N}\right]^{(1 - A_{ij})} \left[\frac{e^{-\frac{1}{\sigma} (N-X_{ij})}}{(1+e^{-\frac{1}{\sigma}})^N}\right]^{ A_{ij}},
\end{align}
which is straightforward to evaluate and simulate numerically.

\begin{table*}
\caption{Variables use in the analysis and algorithm, as defined in terms of the basic quantities $\bm{Z}, \mathcal{D},$ and $\mathcal{A}$.}
\label{table:summary}
\begin{tabular}{cll}
\hline\hline
\null\quad Quantity\quad\null & Definition\hspace{9cm} & Expression\\\hline
$N$ & Number of network samples & $N=\sum_{tu} z_{tu}  $\\
$N_u$ & Number of network samples in cluster for mode~$u$ & $N_u=\sum_{t} z_{tu}$\\
\hline
$M_t$ & Edges in network sample~$t$  & $M_t=\sum_{ij} D_{ij}^{(t)} $\\
$M_u^*$ & Edges in mode~$u$  & $M_u^*=\sum_{ij} A_{ij}^{(u)} $\\
$M^*$ & Edges in all modes  & $M^*=\sum_{iju} A_{ij}^{(u)} $\\
\hline
$X_{ij}^u$ & Number of times $i,j$ are connected in network samples from mode~$u$  & $X_{ij}^u=\sum_t D_{ij}^{(t)} z_{tu}$\\
\hline
$Y_{tu}^{00}$ & Edges absent in sample $t$ and absent in mode $u$ (``true negatives'')  & $Y_{tu}^{00}=\sum_{ij} (1-D_{ij}^{(t)})(1-A_{ij}^{(u)})$\\
$Y_{tu}^{10}$ & Edges present in sample $t$ and absent in mode $u$ (``false positives'') & $Y_{tu}^{10}=\sum_{ij} D_{ij}^{(t)}(1-A_{ij}^{(u)})$\\
$Y_{tu}^{01}$ & Edges absent in sample $t$ and present in mode $u$ (``false negatives'') & $Y_{tu}^{01}=\sum_{ij} (1-D_{ij}^{(t)})A_{ij}^{(u)}$\\
$Y_{tu}^{11}$ & Edges present in sample $t$ and present in mode $u$ (``true positives'') & $Y_{tu}^{11}=\sum_{ij} D_{ij}^{(t)}A_{ij}^{(u)}$\\
\hline
${W}_u^{00}$ & Total true negatives for all samples in cluster for mode $u$  & ${W}_u^{00}=\sum_{ijt} (1-D_{ij}^{(t)})(1-A_{ij}^{(u)}) z_{tu}$\\
${W}_u^{10}$ & Total false positives for all samples in cluster for mode $u$  & ${W}_u^{10}=\sum_{ijt} D_{ij}^{(t)}(1-A_{ij}^{(u)})z_{tu}$\\
${W}_u^{01}$ & Total false negatives for all samples in cluster for mode $u$ & ${W}_u^{01}=\sum_{ijt} (1-D_{ij}^{(t)})A_{ij}^{(u)} z_{tu}$\\
${W}_u^{11}$ & Total true positives for all samples in cluster for mode $u$ & ${W}_u^{11}=\sum_{ijt} D_{ij}^{(t)}A_{ij}^{(u)} z_{tu}$\\
\hline\hline
\end{tabular}
\end{table*}

Comparing with Eq.~\eqref{eq:homogeneous_population_likelihood} of the main text, we see that this model is equivalent to our model with true- and false-positive rates if
\begin{align}
  \frac{e^{-\frac{1}{\sigma} (N-X_{ij})}}{(1+e^{-\frac{1}{\sigma}})^N} &= \alpha^{X_{ij}}(1-\alpha)^{N-X_{ij}},\notag\\
  \frac{e^{-\frac{1}{\sigma} X_{ij}}}{(1+e^{-\frac{1}{\sigma}})^N} &= \beta^{X_{ij}}(1-\beta)^{N-X_{ij}}.
\end{align}
In particular, if we set $\alpha=1-\beta$ then we can map the two models directly, with
\begin{equation}
  \sigma = \left[\log \frac{1-\beta}{\beta}\right]^{-1},
  \label{eq:variance_cer}
\end{equation}
so long as $\beta<1/2$.  In other words, the metric distance approach using the exponential function and Hamming distance~\cite{banks1994metric,lunagomez2020modeling} is equivalent to assuming a binomial model for edge measurements, but with a particular relationship between false-positive and true-positive rates, $\alpha=1-\beta$, which is generally unrealistic~\cite{newman2018estimating,young2020bayesian}.

It is straightforward to verify that analogous results hold in the multimodal case.
One finds that when $\alpha_u = 1 - \beta_u$, one can define $\sigma_u$ analogously to Eq.~\eqref{eq:variance_cer} to obtain a single parameter controlling the probability that existing edges from mode $u$ are missed in the samples and that non-existent edges are added. Thus, we conclude that metric models built with the exponential function and Hamming distance are less expressive than the generative process of network measurements considered here, and we recommend using the latter.  If one does wish to use a metric model then the algorithms presented in the main text can be used to fit it by setting $\alpha_u=1-\beta_u$ for all classes~$u$.

\section{Conjugate priors}
\label{appendix:priors}
For ease of presentation, we have used flat priors for the parameters $\bm{\theta}$ in the main text, but one can easily replace them with more flexible priors thus:
\begin{align}
    P(\alpha_u|H_u^{11}, H_u^{01}) &= \frac{\alpha_u^{H_u^{11} - 1}(1 - \alpha_u)^{H_u^{01} - 1}}{B(H_u^{11}, H_u^{01})}\\
    P(\beta_u|H_u^{00}, H_u^{10}) &= \frac{\beta_u^{H_u^{01} - 1}(1 - \beta_u)^{H_u^{00} - 1}}{B(H_u^{01}, H_u^{00})}\\
    P(\rho|a^*, b^*) &= \frac{\rho^{{a}^* - 1}(1 - \rho)^{b^* - 1}}{B(a^*, b^*)}\\
    P(\pi|\bm{\gamma}) &= \frac{\prod_{u=1}^K \pi_u^{\gamma_k - 1}}{B(\bm{\gamma})}
\end{align}
where $B(x,y)$ is the Euler beta function $B(x,y) = \Gamma(x)\Gamma(y)/\Gamma(x+y)$ and $B(\bm{\gamma})$ is its generalization $B(\bm{\gamma}) = \prod_{u=1}^{K}\Gamma(\gamma_u)/\Gamma(\sum_u \gamma_u) $, and where the parameters of these distributions (such as $H_u^{11}$ and $H_u^{01}$) are new hyper-parameters of the model.

Since the priors above are conjugate~\cite{gelman2013bayesian}, our Gibbs sampling algorithm carries over with little modification.  The only change needed is to the sampling distribution for the parameters:
\begin{align}
  \label{eq:parameter_update_explicit_conjugate}
   P(\bm{\theta}|\mathcal{D},\bm{Z},\mathcal{A})
   \propto{}& \rho^{M^* + a^*}(1-\rho)^{K\binom{N}{2} -M^* + b^*}\notag\\
   &\times \prod_{u=1}^K \pi_u^{N_u + \gamma_u} \notag\\ 
   &\times \prod_{u=1}^K (\alpha_u)^{W_u^{11} + H_u^{11}}(1-\alpha_u)^{W_u^{01} + H_u^{01} }\notag\\
   &\times \prod_{u=1}^K  (\beta_u)^{W_u^{10} + H_u^{10}}(1-\beta_u)^{W_u^{00}+  H_u^{00}}.
\end{align}
The parameters of the conjugate priors are analogous to the combinatorial quantities of Table~\ref{table:summary} and they thus act as pseudo-counts of edges, missing edges, and so on.

\end{document}